%% file: main.tex
\documentclass[
aps,
prl,
superscriptaddress,
citeautoscript,
amsmath,
amssymb,
amsfonts,
showkeys,
floatfix,
reprint,
]{revtex4-2}

\usepackage{xparse}
\usepackage{ifthen}
\newboolean{draft}
\usepackage{xcolor}
\usepackage{amsmath}
\usepackage{mathtools}
\usepackage{amssymb}
\usepackage{amsfonts}
\usepackage{tcolorbox}
\usepackage{xcolor}
\tcbuselibrary{skins,breakable}

\definecolor{ChapterBackground}{HTML}{101010}
\definecolor{ChapterForeground}{HTML}{e93820}
\definecolor{CoverForeground}{HTML}{ee0000}

\newtcolorbox{derivation}[2][]{%
    enhanced,
    skin first = enhanced,
    skin middle = enhanced,
    skin last = enhanced,
    breakable,
    boxrule = 0pt,
    frame hidden,
    borderline west = {4pt}{0pt}{ChapterBackground},
    colback = CoverForeground!10,
    coltitle = ChapterForeground!85,
    sharp corners,
    rounded corners = southeast,
    rounded corners = northeast,
    arc is angular,
    arc = 3mm,
    attach boxed title to top left,
    boxed title style = {%
        enhanced,
        colback = ChapterBackground,
        colframe = ChapterBackground,
        top = 0pt,
        bottom = 0pt,
        sharp corners,
        rounded corners = northeast,
        arc is angular,
        arc = 2mm,
        rightrule = 0pt,
        bottomrule = 0pt,
        toprule = 0pt,
    },
    title = {\bfseries #2},
    overlay unbroken = {%
        \node[anchor=west, color=black!70] at (title.east) {#1};
    },
    overlay first = {%
        \node[anchor=west, color=black!70] at (title.east) {#1};
        \path[fill = tcbcolback!80!black]
            ([yshift = 3mm]interior.south east) -- ++(-0.4,-0.1) -- ++(0.1,-0.2);
    },
    overlay middle = {%
        \path[fill = tcbcolback!80!black]
            ([yshift = -3mm]interior.north east) -- ++(-0.4,0.1) -- ++(0.1,0.2);
        \path[fill = tcbcolback!80!black]
            ([yshift = 3mm]interior.south east) -- ++(-0.4,-0.1) -- ++(0.1,-0.2);
    },
    overlay last = {%
        \path[fill = tcbcolback!80!black]
            ([yshift = -3mm]interior.north east) -- ++(-0.4,0.1) -- ++(0.1,0.2);
    }
}

\setboolean{draft}{True}
\setboolean{draft}{False}

\NewDocumentCommand\change{om}{%
  \ifthenelse{\boolean{draft}}
  {\IfNoValueTF{#1}{{\color{orange}#2}}{{\color{lightgray}#1}/{\color{orange}#2}}}
  {#2}%
}

\def\NV{\text{NV}}%
\def\Ns{\text{N}}
\def\NVm{\text{NV}^{-}}

\def\ri{r}

\newcommand{\braket}[2]{\left\langle#1\middle|#2\right\rangle}
\newcommand{\matrixel}[3]{\left\langle#1\middle|#2\middle|#3\right\rangle}

\begin{document}

\title{Charge state equilibration of nitrogen-vacancy center ensembles in diamond: The role of electron tunneling}

\author{Audrius Alkauskas}
\thanks{Deceased}
\affiliation{Department of Fundamental Research, Center for Physical Sciences and Technology (FTMC), Vilnius LT--10257, Lithuania}

\author{Chris G. Van de Walle}
\affiliation{Materials Department, University of California, Santa Barbara, California 93106-5050, USA}

\author{Lukas Razinkovas}
\email{lukas.razinkovas@ftmc.lt}
\affiliation{Department of Fundamental Research, Center for Physical Sciences and Technology (FTMC), Vilnius LT--10257, Lithuania}

\author{Ronald Ulbricht}
\email{ulbricht@mpip-mainz.mpg.de}
\affiliation{Max Planck Institute for Polymer Research, Ackermannweg 10, 55128 Mainz, Germany}

\begin{abstract}
	The charge state stability of nitrogen-vacancy (NV) centers critically affects
	their application as quantum sensors and qubits. Understanding charge state
	conversion and equilibration is critical not only for NV centers in diamond
	but also for defects and impurities in wide-bandgap materials in general. The
	mechanisms by which these centers change charge state upon optical or
	electronic excitation without the presence of mobile carriers remain unclear,
	potentially affecting the performance of applications ranging from phosphors
	to power electronics. Here, we elucidate this issue for the case of
	photoionization of NV center ensembles. Using pump-probe spectroscopy, we
	ionize negatively charged NV centers and monitor the recovery of $\NVm$ on
	timescales of up to several seconds. We find that the recovery rate depends
	strongly on the concentration of surrounding nitrogen donors. Remarkably, the
	equilibration dynamics exhibit no discernible dependence on temperature,
	ruling out thermally activated processes. The multiphonon-assisted electron
	tunneling model, supported by density-functional calculations, explains the
	measurements and identifies tunneling as the equilibration mechanism.
\end{abstract}

\maketitle

The nitrogen-vacancy ($\NV$) center in diamond~\cite{doherty2013a} has become a
versatile platform for testing and implementing quantum technologies in sensing,
communication, and
computing~\mbox{\cite{schirhagl2014a,hensen2015,awschalom2018a,bradley2019,pezzagna2021}}.
The negatively charged state ($\NVm$) is pivotal for these applications because
it offers the required optical and spin properties. Interactions with nearby
deep-level dopants affect the equilibrium charge state of the NV
center~\cite{collins2002}. To establish the negative charge state, NV typically
captures an electron from a neighboring single-substitutional
nitrogen (N) defect, forming an \mbox{$\NVm$--$\Ns^{+}$} pair~\cite{manson2018}.
Such charge transfer between deep-level defects cannot be explained by the
Shockley-Read-Hall model~\cite{shockley1952,hall1952}, which depends on thermal
excitation and carrier capture. For deep defects with high ionization energies,
this mechanism is ineffective, suggesting that electron tunneling is responsible
for the charge transfer~\cite{manson2018}. Electron tunneling from the excited
state of $\NVm$ to nearby $\Ns^+$ defects has also been proposed as a mechanism
for quenching $\NVm$ photoluminescence (PL)~\cite{luu2024,Capelli2022}. Related
studies of direct inter-defect charge transfer in silicon~\cite{frens1994} and
tunneling through deep levels in quantum-dot heterostructures~\cite{sercel1995}
highlight that tunneling between localized states can be highly efficient.
Despite suggestions, there is a lack of experimental or theoretical evaluations.
Understanding these mechanisms is essential for applications beyond the NV
center, where optical or electronic excitation alters the charge state of deep
defects or impurities in wide-bandgap semiconductors or insulators, such as
phosphors, scintillators, or devices with \mbox{(semi-)}insulating layers.

In this Letter, we investigate the charge equilibration dynamics of an ensemble
of $\NV$ centers following photoionization, which converts $\NV^{-}$ centers to
the neutrally charged $\NV^{0}$ state. Our transient spectroscopy results reveal
that the recovery dynamics back to $\NV^{-}$ are governed by the concentration
of $\Ns$ centers with no discernible dependence on temperature. We propose that
the charge equilibration occurs via phonon-assisted electron tunneling between
$\Ns^{0}$ and $\NV^{0}$, facilitated by significant lattice relaxation of the
$\Ns$ center. This hypothesis is supported by density functional theory (DFT)
calculations of the electron--phonon interaction. We also provide a microscopic
empirical model that replicates the observed dynamics and explains the lack of
temperature dependence. Our approach should be applicable to other materials
containing ensembles of deep-level defects, where electron tunneling serves as
the mechanism for charge migration and equilibration.

\begin{figure}
	\centering
	\includegraphics[width=1\linewidth]{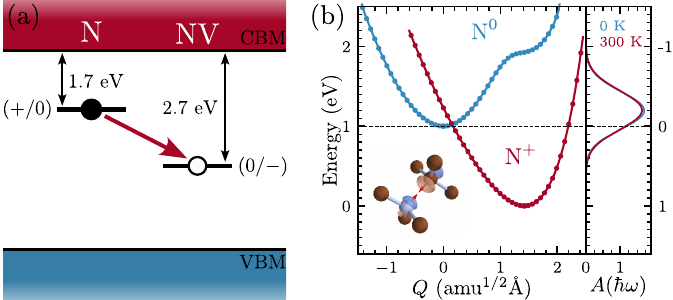}
	\caption{Phonon-assisted electron tunneling from $\Ns^{0}$ to NV$^{0}$.
		(a)~Exothermic charge transfer: the ionization
		threshold for $\Ns^0$ is 1.7~eV, and the energy released when the electron is captured by $\NV^0$ is about 2.7~eV.
		(b)~Left: first-principles configuration coordinate (CC) potential energy
		surfaces of the $\Ns$ center. Inset: atomic structure of N0 with antibonding
		orbital isosurfaces in neutral state; arrows show atomic motion along the
		positive CC $Q$ direction. Right: Optical spectral functions for the
		tunneling process at $T = 0$~K and 300~K, where the value at $E = 0$
		corresponds to the vibrational factor $A(\Delta E; 0)$ in
		Eq.~\eqref{eq:rate}.\label{fig:system}}
\end{figure}

Optical excitation can photoionize $\NVm$ by promoting an electron to the
conduction band (CB), thereby converting $\NVm$ to
$\NV^{0}$~\cite{beha2012,siyushev2013,aslam2013,razinkovas2021a}. The threshold
for direct ionization is estimated to lie between 2.6 and 2.74~eV, based on both
experimental~\cite{aslam2013,bourgeois2017} and
theoretical~\mbox{\cite{deak2014,bourgeois2017,razinkovas2021a}} studies.
Time-resolved THz spectroscopy shows that electrons photoexcited from $\Ns$ to
the CB recombine back to $\Ns$ within tens of
picoseconds~\cite{ulbricht2017c,ulbricht2011a}, indicating a strong
recombination pathway. A similar recombination with $\Ns$ is likely to follow
photoionization of $\NV$. The ionization threshold of $\Ns$ from the neutral to
the positively charged state is approximately
1.7~eV~\cite{nesladek1998,collins2002}. The resulting model of the tunneling
transition that equilibrates the charge after photoionization is depicted in
Fig.~\ref{fig:system}(a), illustrating the charge state transition (CST) levels:
$(0/-)$ for the $\NV$ center and $(+/0)$ for the $\Ns$ defect. The tunneling of
an electron from $\Ns^{0}$ to $\NV^{0}$ is an exothermic process that releases
approximately 1~eV of energy.\looseness=-1

To establish a theoretical foundation for the tunneling process, we consider the
multiphonon emission mechanism, which involves an energy-conserving transition
followed by rapid vibrational relaxation to thermal equilibrium; this relaxation
typically occurs on picosecond timescales~\cite{may2023,ulbricht2018a}. We
employ a simplified microscopic model with three degrees of freedom: one for the
electron and two for the vibrational motion associated with the $\NV$ center and
the $\Ns$ donor, respectively. The charge tunneling rate between two spatially
separated states in the coupled $\NV$--$\Ns$ system is described by Fermi's
golden rule:
\begin{align}
	\label{eq:rate}
	\Gamma = \frac{2\pi}{\hbar} W_{fi}^{2} A(\Delta E;0),
\end{align}
where $W_{fi}$ is the hopping matrix element representing the electronic charge
transfer between the initial ($i$) and final ($f$) electronic states. The
vibrational factor $A(\Delta E,0)$ reflects the probability that the energy
difference between the initial and final electronic states is entirely
transferred to the vibrational system, allowing a non-radiative transition. This
factor corresponds to the value of the normalized spectral function
$A(\Delta E;\hbar\omega)$ at $\hbar\omega=0$, which describes the spectral
density of phonon-assisted radiative
transitions~\cite{henry1977,alkauskas2012a}. In our model, this spectral
function takes the form:
\begin{align}
	\notag
	A(\Delta E, \hbar\omega) =
	 &
	\sum_{mlnk} w_{m}(T) w_{l}(T)
	\left|
	\bigl\langle {\chi_{fn}^{\Ns}}| {\chi_{im}^{\Ns}}\bigr\rangle
	\braket{\chi_{fk}^{\NV}}{\chi_{il}^{\NV}}
	\right|^{2}
	\\
	 &
	\times \delta\left(\Delta E + \varepsilon_{i;ml} - \varepsilon_{f;nk} - \hbar\omega\right),
	\label{eq:lineshape_function}
\end{align}
where $\varepsilon_{i;ml}$ represents the vibrational energy of the system in
the $i$ electronic state, with $\Ns$ and $\NV$ having $m$ and $l$ excited
phonons, respectively. The factors $w_{l}(T)$ are Boltzmann weights that account
for the temperature-dependent occupation of vibrational states.

The vibrational overlap integrals, $\braket{\chi_{fn}}{\chi_{il}}$, quantify the
coupling between vibrational states in the initial and final electronic
manifolds. This electron-phonon interaction is commonly characterized by the
Huang--Rhys (HR) factor $S$~\cite{huang1950}, which represents the average
number of phonons excited during an electronic transition. For the $\NV$ center,
the HR factor for the $(0/-)$ charge-state transition is
$S \approx 1.3$~\cite{razinkovas2021a}, indicating a low probability of exciting
more than two phonons. In contrast, the $\Ns$ defect exhibits much stronger
electron-phonon coupling, with $S \gg 1$~\cite{enckevort1992}, making it the
dominant contributor to the lineshape function. This pronounced coupling arises
from its structural configuration: $\Ns^{0}$ lacks a covalent bond with one
adjacent carbon, forming an antibonding orbital in the region of the missing
bond [see inset of Fig.~\ref{fig:system}(b)]. In its neutral state, this orbital
is occupied, and the removal of the electron restores the covalent bond, leading
to significant lattice relaxation. Given the distinct electron-phonon
interactions, we assume
$\langle \chi^{\NV}_{fn}|\chi^{\NV}_{im}\rangle = \delta_{nm}$, meaning the
$\Ns$ defect solely determines the vibrational factor in
Eq.~\eqref{eq:lineshape_function}. This simplifies the model's functional form
to the well-known 1D problem of carrier capture at point defects
\cite{henry1977,alkauskas2014b}.

We investigate the electron-phonon interaction of the $\Ns$ defect using DFT
calculations within the one-dimensional configurational coordinate (CC)
approximation~\cite{alkauskas2012a}. The calculations were performed using the
VASP code~\cite{kresse1996b} with the r$^{2}$SCAN density
functional~\cite{furness2020}, which has proven effective in describing
electron--phonon interactions in various deep-level
defects~\cite{maciaszek2023,silkinis2024a,silkinis2025,zalandauskas2025a}.
Computational details are available in Sec.~S1 of the Supplemental Material
(SM)~\cite{SM}.

The right side of Fig.~\ref{fig:system}(b) shows the CC potential energy
surfaces for electron tunneling from the neutral $\Ns$ defect to the neutral
$\NV$ center, with a vertical offset of $\Delta E = 1$~eV. Both surfaces exhibit
strong anharmonicity due to pronounced electron--phonon interaction. To
accurately compute the spectral function~\eqref{eq:lineshape_function}, we
numerically solved the one-dimensional nuclear Schrödinger equation. The left
side of Fig.~\ref{fig:system}(b) shows spectral functions
$A(\Delta E; \hbar\omega)$ calculated at $T=0$ and 300~K. The approximate
Huang-Rhys (HR) factor, estimated as the average number of emitted phonons, is
$S \approx 14$. The vibrational factor $A(\Delta E; 0)$ corresponds to the
no-photon-emission condition, and is represented by the horizontal dashed line.
Notably, for $\Delta E = 1$~eV, the factor is near peak efficiency, indicating
an effective channel for non-radiative tunneling. The calculated vibrational
factor values are $A(1, 0) = 1.19~\text{eV}^{-1}$ at $T=0$ and 1.16~eV$^{-1}$ at
300~K, indicating minimal temperature dependence.

For completeness, we also examined an alternative pathway where the $\NVm$
center transitions to the metastable singlet $^{1}\!E$ state, located
${\sim} 0.4$~eV above the ground $^{3}\!A_{2}$ level~\cite{goldman2015}.
However, the vibrational factors for this transition are about 50 times smaller
[$A(0.6, 0) = 0.05~\text{eV}^{-1}$ at $T=0$ and $0.07~\text{eV}^{-1}$ at 300~K],
making this pathway much less significant compared to the transition to the
ground state.

For a qualitative estimate of the electronic transition matrix element $W_{fi}$
and its dependence on distance, we employ the effective-mass approximation and
the zero-range potential (ZRP) model~\cite{bethe1935,lucovsky1965,demkov1988} to
approximate the short-range potential of deep-level defects. In this model, the
wavefunction is expressed as:
\begin{equation}
	\label{eq:psi}
	\psi_{i}(\ri) = \sqrt{\frac{\alpha_{i}}{2\pi}} \frac{e^{-\alpha_{i} \ri}}{\ri},
\end{equation}
where $\ri$ is the radial distance from the defect center. Within the
effective-mass approximation, the parameter $\alpha_{i} = \sqrt{2m E_{i}}/\hbar$
is determined by the ionization energy $E_{i}$. In the ZRP model, the
short-range potential is replaced by a boundary condition requiring that the
logarithmic derivative of the radial part $\ri\psi_{i}$ equals $-\alpha_{i}$.

The electron transfer matrix element, analogous to the hopping term in a
tight-binding model, can be expressed as
$W_{fi} = E_{\NV}\langle \psi_{\NV}|\psi_{\Ns}\rangle + \langle \psi_{\NV}|V_{\NV}|\psi_{\Ns}\rangle$~\cite{sercel1995}.
Here, $E_{\NV}$ is the negative binding energy of the NV center electron,
$\psi_{\NV}$ and $\psi_{\Ns}$ are the localized wavefunctions of the isolated
$\NV$ and $\Ns$ defects, and $V_{\NV}$ is the potential of the $\NV$ center.
Within the ZRP model, the matrix element is derived as follows (see Sec.S2 of
the SM~\cite{SM} for details):
\begin{equation}
	\label{eq:W}
	W_{fi} = \frac{\hbar^{2}}{mr}
	\frac{\sqrt{\alpha_{1}\alpha_{2}}}{\left(\alpha_{1}^{2}-\alpha_{2}^{2}\right)}
	\left(\alpha_{2}^{2}e^{-\alpha_{1}r} - \alpha_{1}^{2}e^{-\alpha_{2}r}\right),
\end{equation}
where $r$ is the separation between the centers.

To model the population dynamics of ionized $\NV$ centers, we assume that the
electron tunnels from the nearest $\Ns^{0}$ defect. This assumption is justified
by the experimental protocol, which involves repeated intense optical pulses
that ionize the $\NVm$ centers and redistribute the local charge. As a result,
the probability of finding a nearby positively charged nitrogen ($\Ns^{+}$)
immediately after excitation is significantly reduced. We therefore assume that
nearly all substitutional nitrogen remains in the neutral charge state. The
radial probability distribution of finding the nearest neighbor in the radial
shell between $r$ and $r+\mathrm{d}r$ is given by~\cite{Chandrasekhar1943}:
\[
	w(r)=4\pi r^{2}ne^{-\frac{4}{3}\pi n r^{3}}.
\]
where $n$ is the overall density of nitrogen defects. The function $w(r)$ peaks
at a distance of $(2\pi n)^{-1/3}$, corresponding to 44.9, 20.9, and 17.1~\AA\
for nitrogen concentrations of 10, 100, and 180 ppm, respectively.

Using the above distribution of nearest $\Ns$ donors, the temporal dependence of
the relative concentration of neutral $\NV$ centers,
$p(t) = [\NV^{0}](t)/[\NV^{0}](0)$, due to tunneling from $\Ns$ defects,
becomes:
\begin{equation}
	\label{eq:p0}
	p(T, t) = \int_{0}^{\infty} w(r) e^{-\Gamma(r;T)t}\,\mathrm{d}r.
\end{equation}

\begin{figure}
	\centering
	\includegraphics[width=0.98\linewidth]{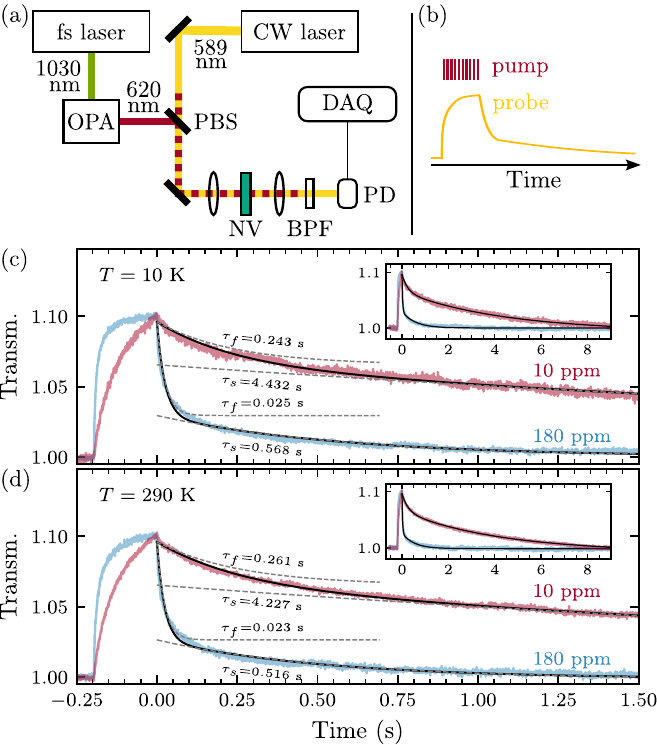}
	\caption{(a) Schematic of the measurement setup. Abbreviations: OPA -- optical
		parametric amplifier, PBS -- polarizing beam splitter, BPF -- bandpass
		filter, PD -- photodiode, DAQ -- Data acquisition card; (b) Excitation
		scheme illustrating pump and probe dynamics; Normalized probe transmission
		dynamics for both samples at temperatures of c) 10~K and d) 290~K, with
		bi-exponential fits to the data. The inset plots display
		graphs for a 10~s time window.\label{fig:setup}}
\end{figure}

Having established a theoretical framework for modeling charge equilibration
dynamics via electron tunneling, we now turn to experimental verification. We
employ a pump–probe spectroscopy setup using two diamond substrates having low
(10~ppm) and high (180~ppm) $\Ns$ concentrations. A pulsed pump laser ionizes
the NV ensemble, while a continuous-wave (CW) laser monitors the time-resolved
transmission through the sample~\cite{Younesi2022}. The setup is illustrated in
Fig.~\ref{fig:setup}(a), with additional details provided in Sec.~S4 of the
SM~\cite{SM}.

In brief, pump pulses with a duration of approximately 150~fs at a wavelength of
620~nm are emitted at a 10~kHz repetition rate. For the considered timescales,
the pulsed excitation can be treated as a continuous excitation. The chosen
wavelength selectively excites and, via resonant two-photon
absorption~\cite{razinkovas2021a}, ionizes $\NV^{-}$ without exciting $\NV^{0}$.
The CW probe laser emits at a wavelength of 589~nm, again selectively addressing
$\NV^{-}$ without exciting $\NV^{0}$. The time-resolved measurement of the
transmitted probe beam enables monitoring of the transient bleaching of the
$\NV^{-}$ absorption. Figures~\ref{fig:setup}(c) and (d) show the normalized
probe transmission dynamics for both samples at 10 and 290~K, respectively, with
the pre-pump signal set to unity and the maximum set to 1.1. The pump laser
excites the sample for 200~ms, steadily building a non-equilibrium concentration
of $\NV^0$ centers, as evidenced by an increase in the probe transmission, i.e.,
bleaching of $\NV^{-}$ absorption. After switching off the pump, the system
relaxes towards equilibrium for up to 10 seconds while the probe laser monitors
the process. This cycle is repeated several times, and the recorded transient
transmission data are averaged.

The relaxation dynamics can be approximately described by a bi-exponential
model, as indicated by the fits shown in Figs.~\ref{fig:setup}(c) and (d), where
the corresponding characteristic time constants are displayed. Notably, the
dynamics remain nearly identical at both cryogenic and room temperatures. A more
quantitative analysis is provided in Sec.~S4 of the SM~\cite{SM}, where the
extracted time constants are plotted as a function of temperature for both
samples, confirming the temperature independence predicted by our model.

To relate the theoretical model to the experimental conditions, we now consider
the population buildup during the pumping stage. Applying ionizing excitation
results in the accumulation of a population of $\NV^0$ centers with longer
equilibration times, while effectively filtering out contributions with
tunneling rates higher than the average ionization rate. Accordingly,
Eq.~\eqref{eq:p0} is adjusted as:
\begin{equation}
	\label{eq:filtered}
	p_{\text{eff}}(T, t) = \int_{0}^{\infty} f\left(\tau[r;T]\right) w(r)  e^{-\Gamma(r;T) t} \,\mathrm{d}r,
\end{equation}
where $f(\tau)$ is a filter function that describes how the population
redistributes due to the pumping sequence, and $\tau = 1/\Gamma(r;T)$ is the
tunneling lifetime, which depends on the distance between NV and $\Ns$ centers.
The function $f(\tau)$ can be estimated by treating the pulsed excitation as
continuous, replacing the pump sequence with a constant average rate~$G$. In the
weak-excitation limit ($G \ll \Gamma$), the steady-state population after a pump
duration $\Delta t$ takes the form (see Sec.~S3 of the SM for details):
\begin{equation}
	\label{eq:3}
	f(\tau) = G \tau \left(1 - e^{-\Delta t / \tau} \right).
\end{equation}

\begin{figure}
	\centering
	\includegraphics[width=1\linewidth]{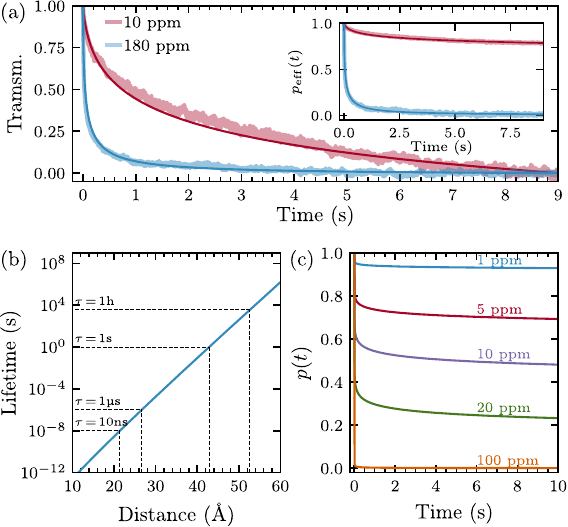}
	\caption{(a)~Simultaneous fit of the empirical tunneling model (thin solid
		lines) to the experimental decay curves at $T=10$~K (thick blurry lines) for
		10~ppm and 180~ppm nitrogen concentrations. Both datasets are normalized
		such that the pre-pump population is one and the value at 9~s is zero.
		Inset: Same data plotted on an absolute scale, with experimental curves
		rescaled to match the predicted population difference. (b)~Calculated
		tunneling lifetime $\tau = 1/\Gamma$ as a function of $\NV$--$\Ns$
		separation at $T = 0$~K. (c)~Time evolution of the relative $\NV^{0}$
		population $p(t)$ at $T = 0$~K for different $\Ns$ concentrations
		(1--100~ppm), following a single-pulse ionization.\label{fig:theory2}}
\end{figure}

The empirical model introduced above relies on two phenomenological parameters,
$\alpha_{1}$ and $\alpha_{2}$, which describe the spatial extent of the $\Ns$
and $\NV$ wavefunctions, respectively. Assuming an effective electron mass of
$m = 0.57\,m_{e}$, corresponding to the average conduction-band mass in diamond,
and using a pump duration of $\Delta t = 0.2$~s, we simultaneously fit these
parameters to the experimental decay curves at $T=10$~K measured for both
nitrogen concentrations. To enable direct comparison with the experiment, the
simulated and experimental decay curves are normalized such that the pre-pump
population is set to unity and the minimum value at $t = 9$~s is set to 0.
Figure~\ref{fig:theory2}(a) shows the fitted curve overlaid on the measured
data. The inset presents both simulated and experimental curves on an absolute
population scale, with the experimental data rescaled so that the population
change over the 9-second interval matches the prediction of the theoretical fit.
The model accurately captures the qualitative shape of the probe transmission
and reproduces the overall trend observed in both nitrogen concentration
samples, highlighting the difference between the two populations. The fitted
parameters are $\alpha_{1} = 0.40~\text{\AA}^{-1}$ and
$\alpha_{2} = 0.89~\text{\AA}^{-1}$. These values are comparable to those
predicted by the phenomenological effective mass expression
$\alpha_{i} = \sqrt{2mE_{i}}/\hbar$, which yields $0.51~\text{\AA}^{-1}$ and
$0.64~\text{\AA}^{-1}$ for the respective ionization energies.

While the extracted parameters provide only a rough
\change[approximation]{estimate}, we assume that they capture the
\change[essential features]{core aspects} of the tunneling process and use them
to \change[estimate]{determine} the associated relaxation lifetimes.
Figure~\ref{fig:theory2}(b) shows the calculated tunneling lifetime
$\tau = 1/\Gamma$ as a function of $\NV$--$\Ns$ separation at $T = 0$~K. The
lifetime increases rapidly with distance, reflecting the strong exponential
sensitivity of the tunneling process. According to our model,
tunneling occurs on the nanosecond timescale when $\Ns$ donors lie within
20~\AA\ of an $\NV$ center but reaches hours if the separation is above 50~\AA.
Such rapid decay at $\NV$--$\Ns$ separations up to 20~\AA\ qualitatively
\change[agrees]{aligns} with the estimate in Ref.~\cite{manson2018}, which
suggests that $\NVm$ centers appear dark if a positively charged $\Ns$ defect is
\change[located ]{}within 12~\AA, due to electron tunneling from the excited
state of $\NVm$ to a nearby $\Ns^{+}$ center. Although the underlying process
\change[is different]{differs}, the energy released by tunneling from $\NV^{-}$
to $\Ns^{+}$ is also on the order of 1~eV, and the electron--phonon interaction
strength is expected to be of similar magnitude due to the involvement of a
charge state change of the $\Ns$ defect.

Figure~\ref{fig:theory2}(c) \change[shows]{displays} the theoretical
\change[prediction for the time evolution of the relative fraction of $\NV^{0}$
centers]{prediction for how the relative fraction of $\NV^{0}$ centers evolves
  over time} at $T = 0$~K for $\Ns$ donor concentrations of 1, 5, 10, 20, and
100~ppm, assuming a single short-pulse ionization [Eq.~\eqref{eq:p0}]. On a
linear scale, the decay dynamics exhibit two distinct regimes: an initial rapid
decrease in $p(t)$, followed by a much slower decay, a characteristic behavior
also observed in our experimental measurements with a finite pre-pump time. For
the nitrogen concentrations considered here, the slow-decay regime can extend
over several seconds. For example, at 10~ppm, the model predicts that $p(t)$
remains above 0.48 even after ten seconds. At higher nitrogen concentrations,
charge equilibration \change[becomes substantially faster]{happens much faster}:
at 20~ppm, the population drops to 0.23 after 10~s, while at 100~ppm, it falls
below 1\% ($p(t)=0.001$) in less than 100~ms. At lower nitrogen concentrations,
charge equilibration slows \change[considerably]{down significantly}. For the
1~ppm case, our model predicts that only 11\% of the population \change[is
recovered]{recovers} after one hour ($p(t) = 0.89$).

Our combined experimental and theoretical analysis identifies electron tunneling
between N donors and NV centers as the \change[dominant]{primary} mechanism
governing charge-state equilibration in ensembles. The results provide a
quantitative basis for predicting NV charge stability and establish an
experimental protocol for directly probing defect–defect charge transfer. Future
\change[work could examine]{research could explore} how surface-induced band
bending alters defect energetics and tunneling rates, \change[particularly for
sensing applications in which NV centers are intentionally placed near the
surface]{particularly in sensing applications where NV centers are deliberately
  positioned near the surface}. \change[A complete picture of NV charge
stability will also require understanding how other deep dopants in diamond
influence these dynamics, and how charge transfer between dopants (e.g.,
migration of positive charge among nitrogen centers) modifies the local charge
landscape]{Achieving a comprehensive understanding of NV charge stability will
  also involve analyzing how other deep dopants in diamond influence these
  dynamics and how charge transfer among dopants (such as the movement of
  positive charge among nitrogen centers) alters the local charge environment}.
Our framework should also \change[apply]{be applicable } to other deep-level
defects in wide-bandgap materials, \change[i.e.]{especially} in regimes where
carrier interaction with the bands is
suppressed and tunneling governs charge migration.

\textit{Acknowlegments--}We thank Neil Manson for many helpful discussions and
Vytautas Karpus for introducing us to the zero-range potential model. C. G. VdW. thanks C. Vishwakarma and J. K. Nangoi for useful discussions. R.U.\
acknowledges support from the Max Planck Society. A.A.\ and L.R.\ were supported by the QuantERA grant SensExtreme, funded by the Lithuanian Research Council (Grant No.~S-QUANTERA-22-1). Computational resources were provided by the supercomputer GALAX at the Center for Physical Sciences and Technology, Lithuania, and by the High Performance Computing Center “HPC Saulėtekis” at the Faculty of Physics, Vilnius University. C. G. VdW. was supported by the U.S. Department of Energy, Office of Science, Co-design Center for Quantum Advantage (C2QA) under contract number DE-SC0012704.

\textit{Data availability--}The data that support the findings of this study are
available from the corresponding authors upon reasonable request.

\bibliography{references}

\clearpage
\onecolumngrid

\setcounter{section}{0}
\setcounter{figure}{0}
\setcounter{table}{0}
\setcounter{equation}{0}

\setcounter{secnumdepth}{2}   
\renewcommand{\thesection}{S\arabic{section}}
\renewcommand{\thesubsection}{S\arabic{section}.\Alph{subsection}}
\renewcommand{\theequation}{S\arabic{section}.\arabic{equation}}
\renewcommand{\thefigure}{S\arabic{figure}}
\renewcommand{\thetable}{S\arabic{table}}

\section*{Supplemental Material}

\input{sup/ab_initio.tex}

\input{sup/transfer_integral.tex}

\input{sup/pumping.tex}

\input{sup/ensemble.tex}

\input{sup/exp.tex}

\end{document}

%% file: sup/ab_initio.tex
\section{First-Principles Modeling of Electron-Phonon Interaction}
\label{sec:ab_initio}

We model the electron-phonon interaction of the $\Ns$ defect using a
one-dimensional configurational coordinate (CC) approach~\cite{alkauskas2012a},
which is well-suited for systems with strong electron-phonon coupling, as is the
case in this study. In this approach, the complex multi-mode vibrational motion
is approximated by a single degree of freedom, the configurational coordinate
$Q$, which parametrizes the ionic displacement between equilibrium geometries of
the defect in two charge states. The total displacement between two geometries,
$\Delta Q$, in mass-weighted coordinates is given by $\Delta Q^2 = \sum_{\alpha,
  i} M_\alpha \Delta R_{\alpha,i}^2$, where the sum runs over all atoms $\alpha$
and Cartesian directions $i$. Here, $M_\alpha$ is the atomic
mass of species $\alpha$, and $\Delta R_{\alpha, i}$ represents the difference
in equilibrium positions between the two charge states. The potential energy
surfaces for both charge states are then computed along this parametrized
configurational coordinate path.

The calculations were performed using spin-polarized density functional theory
(DFT) with exchange and correlation described by the r$^2$SCAN
functional~\cite{furness2020}. We employed the projector-augmented wave (PAW)
method as implemented in the VASP code~\cite{kresse1996b}, using a plane-wave
energy cutoff of 600~eV. The defect was modeled in a $4 \times 4 \times 4$
face-centered cubic (fcc) supercell containing 512 atoms, with the Brillouin
zone sampled at a single $\Gamma$ point. The equilibrium geometries were
optimized until the Feynman–Hellmann forces on each atom were below
0.001~eV$/$\AA.

\begin{figure*}[b]
  \centering
  \includegraphics[width=0.8\linewidth]{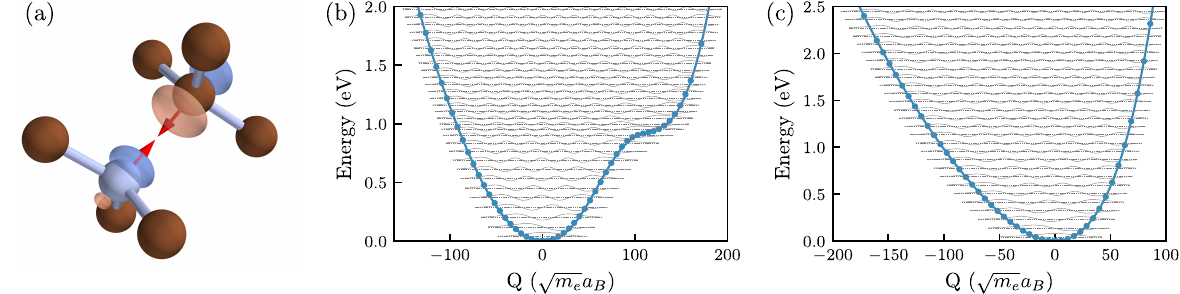}
  \caption{Ab initio characterization of the electron--phonon interaction in the
    $\Ns$ defect. (a) Atomic structure of the neutral $\Ns$ center, with
    isosurfaces showing the antibonding localized orbital occupied in the
    neutral state; arrows indicate atomic motion along the positive CC $Q$
    direction. (b) and (c) Configuration coordinate (CC) potential energy
    surfaces for $\Ns^{0}$ and $\Ns^{+}$, with dots representing DFT
    calculations. Numerical solutions of nuclear wavefunctions are shown with
    baselines shifted according to vibrational state energies.\label{fig:ep}}
\end{figure*}

In Fig.~\ref{fig:ep}(a), we present the geometry of the $\Ns$ defect in its
neutral charge state. The absence of a covalent bond with one adjacent carbon
reduces the symmetry from $T_d$ to $C_{3v}$. Instead of a covalent bond, a
singly occupied antibonding orbital is oriented along the $C_{3v}$ axis,
representing the only $\Ns$-derived state within the diamond band gap. The
nitrogen-carbon bond length to the three nearest neighbors is 1.47~\AA, while
the bond length along the symmetry axis is 2.03~\AA, highlighting significant
symmetry breaking. Upon ionization to the positively charged state, electron
removal restores $T_d$ symmetry, with all nitrogen-carbon bond lengths
equalized at 1.56~\AA. This substantial structural relaxation along the
symmetry axis of $\Ns^{0}$ leads to strong electron--phonon interaction.

Figures~\ref{fig:ep}(b) and (c) present the potential energy surfaces computed
along the configurational coordinate. Both surfaces exhibit strong anharmonicity,
making the harmonic approximation inadequate for describing electron--phonon
interactions. To accurately compute the vibrational overlap integrals entering
the normalized spectral function [see also Eq.~(2) of the main text]:
\begin{align}
  A(\Delta E, \hbar\omega) =
  &
    \sum_{mlnk} w_{m}(T)
    \left|
    \bigl\langle {\chi_{fn}}| {\chi_{im}}\bigr\rangle
    \right|^{2}
    \times \delta\left(\Delta E + \varepsilon_{i;ml} - \varepsilon_{f;nk} - \hbar\omega\right),
       \label{eq:spectral_function}
\end{align}
we numerically solve for the vibrational states of both potentials. The Numerov
integration scheme combined with a shooting method is used to determine the 30
lowest bound states for each surface. The lowest-lying states are visualized in
Figs.~\ref{fig:ep}(b) and (c). These computed states form the basis for
evaluating the spectral function~\eqref{eq:spectral_function}. To achieve a
smooth representation of the multi-mode system [as shown on the right side of
Fig.~1(b)], the delta functions are replaced with Gaussians of width
$\sigma = 60$~meV.


%% file: sup/transfer_integral.tex
\global\long\def\rb{\boldsymbol{r}}%
\global\long\def\Rb{\boldsymbol{R}}%
\def\dd{\mathrm{d}}

\section{Zero-range potential (ZRP) model for electron tunneling between two deep-level defects}
\label{sec:ZRP}

In this section, we develop an analytical expression for the electron tunneling
transfer integral between two deep-level defects, focusing on the limit where
the binding well width approaches a point-like limit.

\subsection{Derivation of the Transfer Integral}

An electron localized on a deep-level defect site experiences a short-range
potential that decays rapidly with distance from the defect. Beyond a
certain cutoff distance, we assume that the wave function behaves as that of a
free particle. The simplest model to capture this behavior is the zero-range
potential (ZRP) model, originally proposed by Bethe and
Peierls~\cite{bethe1935}. In the ZRP model, the wave function beyond the
point-like potential is that of a free particle:
\begin{equation}
	\label{eq:spsi}
	\psi_{i}(\ri) = c_{i} \frac{e^{-\alpha_{i} \ri}}{\ri},
\end{equation}
where $\ri = \lvert \rb - \Rb_{i} \rvert$ denotes the radial distance from the
defect located at $\Rb_{i}$, and $c_{i}=\sqrt{{\alpha_{i}}/{2\pi}}$ serves as the
normalization constant. The parameter $\alpha_{i} = \sqrt{2m(-E_{i})}/\hbar$
defines the depth of the potential well, with $-E_{i}$ representing the binding
(ionization) energy and $m$ the effective electron mass. Instead of explicitly
defining the potential, the ZRP model replaces it with a boundary condition on
the logarithmic derivative of the wave function:
\[
	\left. \frac{1}{\ri \psi_{i}} \frac{\partial (\ri \psi_{i})}{\partial \ri}
	\right|_{\ri \to 0} = -\alpha_{i}.
\]
The action of the Hamiltonian describing this potential on its eigenfunction
$\psi_{i}$ is given by a Green's function equation~\cite{demkov1988}:
\begin{align}
	\left(-\frac{\hbar^{2}}{2m} \nabla^{2} - E_{i}\right) \psi_{i}(\rb)
	 & = \frac{2\pi\hbar^{2}}{m} c_{i} \delta(\rb - \Rb_{i}). \label{eq:green}
\end{align}
This formulation allows the potential well to be expressed as:
\begin{equation}
	\label{eq:ZRPpot}
	V_{i} = -\frac{2\pi\hbar^{2}}{m} c_{i} \delta(\rb - \Rb_{i}) \frac{1}{\psi_{i}}.
\end{equation}

In the basis of two localized wave functions, $\psi_{1}$ and $\psi_{2}$, which
are the ground-state eigenfunctions of isolated ZRP Hamiltonians with point-like
potential wells separated by a distance $r$, the zero-order Hamiltonian $H_{0}$
describes a system in which each wave function interacts solely with its own local
potential, with no influence from the neighboring potential. The Hamiltonian is
expressed as:
\[
	H_{0} = \begin{pmatrix}
		E_{1} & 0     \\
		0     & E_{2}
	\end{pmatrix}.
\]

When two centers are in proximity, the full Hamiltonian $H = T + V_{1} + V_{2}$
accounts for the potentials of both ZRP wells, $V_{1}$ and $V_{2}$. The matrix
elements of the full Hamiltonian, in the basis of the wavefunctions of the
isolated system, are given by:
\begin{subequations}
	\begin{align}
		\matrixel{\psi_{1}}{H}{\psi_{1}} & = E_{1} + \matrixel{\psi_{1}}{V_{2}}{\psi_{1}},                    \\
		\matrixel{\psi_{2}}{H}{\psi_{2}} & = E_{2} + \matrixel{\psi_{2}}{V_{1}}{\psi_{2}},                    \\
		\matrixel{\psi_{1}}{H}{\psi_{2}} & = \matrixel{\psi_{1}}{T + V_{2}}{\psi_{2}}
		+ \matrixel{\psi_{1}}{V_{1}}{\psi_{2}}\notag                                                          \\
		                                 & = -\frac{\alpha_{2}^{2} \hbar^{2}}{2m} \braket{\psi_{1}}{\psi_{2}}
		- \frac{\sqrt{\alpha_{1} \alpha_{2}} \hbar^{2}}{r m} e^{-\alpha_{2} r}.
		\label{eq:ta1}
	\end{align}
\end{subequations}
Equation~\eqref{eq:ta1} is derived using the relation
$E_{2} = -\alpha_{2}^{2} \hbar^{2} / 2m$ and
$\matrixel{\psi_{1}}{V_{1}}{\psi_{2}} = -(2\pi \hbar^{2} / m) c_{1} \psi_{2}(\Rb_{1})$.
The overlap integral between two wavefunctions of the form~\eqref{eq:spsi},
separated by a distance $r$, is given by:
\begin{equation}
  \label{eq:overlap}
  \braket{\psi_{2}}{\psi_{1}} = -\frac{2}{r}
  \frac{\sqrt{\alpha_{1} \alpha_{2}}}{\alpha_{1}^{2} - \alpha_{2}^{2}}
  \left( e^{-\alpha_{1} r} - e^{-\alpha_{2} r} \right).
\end{equation}

\begin{derivation}{Overlap derivation}

	To estimate the overlap integral above, we position
	the first potential well at the origin, $\Rb_{1}=0$, and the second
	at $\Rb_{2}=r\hat{e}_{z}$, where $r$ is the separation distance
	between the two defects. In spherical coordinates, the wave functions
	are defined as follows:
	\begin{align*}
		\psi_{1}(\rho,\theta,\phi) & =c_{1}\frac{e^{-\alpha_{1}\rho}}{\rho},                                                                       \\
		\psi_{2}(\rho,\theta,\phi) & =c_{2}\frac{e^{-\alpha_{2}\sqrt{\rho^{2}+r^{2}-2\rho r\cos\theta}}}{\sqrt{\rho^{2}+r^{2}-2\rho r\cos\theta}}.
	\end{align*}

	The overlap integral $\langle\psi_{2}|\psi_{1}\rangle$ then takes the form:
	\begin{align*}
		\braket{\psi_{2}}{\psi_{1}}
		 & =
		c_{1}c_{2}\int_{0}^{2\pi}\dd\phi\int_{0}^{\pi}\dd\theta\sin\theta\int_{0}^{\infty}\dd\rho\,
		\rho^{2}\frac{e^{-\alpha_{1}\rho}}{\rho}\frac{e^{-\alpha_{2}\sqrt{\rho^{2}+r^{2}-2\rho r \cos\theta}}}{\sqrt{\rho^{2}+r^{2}-2\rho r \cos\theta}}
		\\
		 &
		=2\pi c_{1}c_{2}\int_{0}^{\pi}\dd\theta\sin\theta\int_{0}^{\infty}\dd\rho\,\rho\frac{e^{-\alpha_{1}\rho-\alpha_{2}\sqrt{\rho^{2}+r^{2}-2\rho r \cos\theta}}}{\sqrt{\rho^{2}+r^{2}-2\rho r \cos\theta}},
	\end{align*}

	By substituting $u=\cos\theta$ and $\dd u=-\sin\theta\:\dd\theta$,
	the integral simplifies to:
	\begin{align*}
		\braket{\psi_{2}}{\psi_{1}}
		 & =2\pi c_{1}c_{2}\int_{0}^{\infty}\dd\rho\,\int_{-1}^{1}\dd u\,\rho\frac{e^{-\alpha_{2}\sqrt{\rho^{2}+r^{2}-2\rho r  u}-\alpha_{1}\rho}}{\sqrt{\rho^{2}+r^{2}-2\rho r  u}}                            \\
		 & =2\pi c_{1}c_{2}\int_{0}^{\infty}\dd\rho\,\underbrace{\int_{-1}^{1}\dd u\,e^{-\alpha_{1}\rho}\frac{\rho e^{-\alpha_{2}\sqrt{\rho^{2}+r^{2}-2\rho r  u}}}{\sqrt{\rho^{2}+r^{2}-2\rho r  u}}}_{I_{u}}.
	\end{align*}

	To estimate $I_{u}$, we make a further substitution:
	\[
		v=e^{-\alpha_{2}\sqrt{r^{2}-2d\rho u+\rho^{2}}},\quad\text{with }\dd v=\alpha_{2} r \frac{\rho e^{-\alpha_{2}\sqrt{r^{2}-2d\rho u+r^{2}}}}{\sqrt{r^{2}-2d\rho u+\rho^{2}}}\dd u,
	\]
	which transforms the integral to:
	\begin{align*}
		I_{u}
		 & =\frac{1}{\alpha_{2} r }e^{-\alpha_{1}\rho}\int_{v(-1)}^{v(1)}\dd v
		=\frac{1}{\alpha_{2} r }\left(\underbrace{\exp\left[-\alpha_{1}\rho-\alpha_{2}\sqrt{( r -\rho)^{2}}\right]}_{I_{u_{1}}}-\underbrace{\exp\left[-\alpha_{1}\rho-\alpha_{2}\sqrt{( r +\rho)^{2}}\right]}_{I_{u_{2}}}\right).
	\end{align*}

	We first estimate the integral $I_{1}=\int_{0}^{\infty}\dd \rho\,I_{u_{1}}$
	by splitting it into two ranges:
	\begin{align*}
		I_{1}=
		  & \int_{0}^{\infty}\dd \rho\,\exp\left[-\alpha_{1}\rho-\alpha_{2}\sqrt{( r -\rho)^{2}}\right]                                                                                   \\
		= & \int_{0}^{ r }\dd \rho\,\exp\left[-(\alpha_{1}-\alpha_{2})\rho-\alpha_{2} r \right]+\int_{ r }^{\infty}\dd \rho\,\exp\left[-(\alpha_{1}+\alpha_{2})\rho+\alpha_{2} r )\right] \\
		= & -\frac{e^{-\alpha_{1} r }-e^{-\alpha_{2} r }}{\alpha_{1}-\alpha_{2}}+\frac{e^{-\alpha_{1} r }}{\alpha_{1}+\alpha_{2}}.
	\end{align*}
	The integral $I_{2}=\int_{0}^{\infty}\dd \rho\,I_{u_{2}}$ estimates
	to
	\[
		I_{2}=\int_{0}^{\infty}\dd \rho\,\exp\left[-\alpha_{1}\rho-\alpha_{2}( r +\rho)\right]=\frac{e^{-\alpha_{2} r }}{\alpha_{1}+\alpha_{2}}.
	\]
	The final expression for the overlap integral is estimated as follows:
	\begin{align*}
		\braket{\psi_{2}}{\psi_{1}}
		 & =\frac{2\pi c_{1}c_{2}}{\alpha_{2} r }\left[-\frac{e^{-\alpha_{1} r }-e^{-\alpha_{2} r }}{\alpha_{1}-\alpha_{2}}+\frac{e^{-\alpha_{1} r }}{\alpha_{1}+\alpha_{2}}-\frac{e^{-\alpha_{2} r }}{\alpha_{1}+\alpha_{2}}\right]
		\\
		 & =-\frac{2\pi c_{1}c_{2}}{\alpha_{2} r }\frac{2\alpha_{2}\left(e^{-\alpha_{1} r }-e^{-\alpha_{2} r }\right)}{\alpha_{1}^{2}-\alpha_{2}^{2}}                                                                                \\
		 & =-\frac{2}{ r }\frac{\sqrt{\alpha_{1}\alpha_{2}}}{\left(\alpha_{1}^{2}-\alpha_{2}^{2}\right)}\left(e^{-\alpha_{1} r }-e^{-\alpha_{2} r }\right)
	\end{align*}
	In the case when both wells are identical $\alpha_{2}=\alpha_{1}=\alpha$
	the integral simplifies to:
	\[
		\left.\braket{\psi_{2}}{\psi_{1}}\right|_{\alpha=\alpha_{1}=\alpha_{2}}=e^{- r \alpha}.
	\]

\end{derivation}

Using the equation above, the transfer integral becomes
\begin{equation}
  \label{eq:Wfinal}
  W=\matrixel{\psi_{1}}H{\psi_{2}}= \frac{\hbar^{2}}{mr}\frac{\sqrt{\alpha_{1}\alpha_{2}}}{\left(\alpha_{1}^{2}-\alpha_{2}^{2}\right)}\left(\alpha_{2}^{2}e^{-\alpha_{1}r}-\alpha_{1}^{2}e^{-\alpha_{2}r}\right)
\end{equation}

\begin{derivation}{Derivation}
  \begin{align*}
\matrixel{\psi_{1}}H{\psi_{2}} & =-\frac{\alpha_{2}^{2}\hbar^{2}}{2m}\braket{\psi_{1}}{\psi_{2}}-\frac{\sqrt{\alpha_{1}\alpha_{2}}\hbar^{2}}{mr}e^{-\alpha_{2}r}\\
 & =\frac{\hbar^{2}}{mr}\left(\frac{\alpha_{2}^{2}\sqrt{\alpha_{1}\alpha_{2}}\left(e^{-\alpha_{1}r}-e^{-\alpha_{2}r}\right)}{\left(\alpha_{1}^{2}-\alpha_{2}^{2}\right)}-\sqrt{\alpha_{1}\alpha_{2}}e^{-\alpha_{2}r}\right)\\
 & =\frac{\hbar^{2}}{mr}\frac{\sqrt{\alpha_{1}\alpha_{2}}}{\left(\alpha_{1}^{2}-\alpha_{2}^{2}\right)}\left[\alpha_{2}^{2}\left(e^{-\alpha_{1}r}-e^{-\alpha_{2}r}\right)-\left(\alpha_{1}^{2}-\alpha_{2}^{2}\right)e^{-\alpha_{2}r}\right]\\
 & =\frac{\hbar^{2}}{mr}\frac{\sqrt{\alpha_{1}\alpha_{2}}}{\left(\alpha_{1}^{2}-\alpha_{2}^{2}\right)}\left[\alpha_{2}^{2}e^{-\alpha_{1}r}-\alpha_{2}^{2}e^{-\alpha_{2}r}-\alpha_{1}^{2}e^{-\alpha_{2}r}+\alpha_{2}^{2}e^{-\alpha_{2}r}\right]\\
 & =\frac{\hbar^{2}}{mr}\frac{\sqrt{\alpha_{1}\alpha_{2}}}{\left(\alpha_{1}^{2}-\alpha_{2}^{2}\right)}\left(\alpha_{2}^{2}e^{-\alpha_{1}r}-\alpha_{1}^{2}e^{-\alpha_{2}r}\right)
\end{align*}
\end{derivation}

\subsection{Approximation of Zero Wavefunction Overlap}

\begin{figure}
\centering
\includegraphics[width=0.5\textwidth]{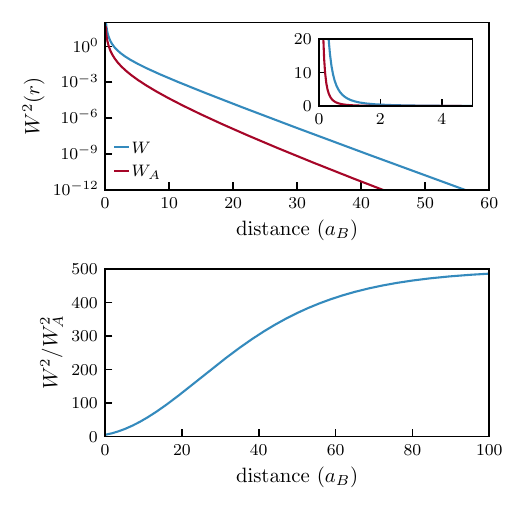}
\caption{Comparison of $W^{2}$ calculated using the full treatment (including overlaps)
and the zero-overlap approximation. The bottom panel shows the ratio of the two
models.\label{fig:WWacomp}}
\end{figure}

In the model described above, we introduce an additional approximation that neglects
the overlap between the wavefunctions. Under this assumption, the
transfer integral simplifies to:
\begin{align}
  W_{A} & = \frac{1}{2} \matrixel{\psi_{1}}{T + V_{1} + T + V_{2} + V_{1} + V_{2}}{\psi_{2}}
		  \notag\\
		& = \frac{1}{2} \matrixel{\psi_{1}}{V_{1} + V_{2}}{\psi_{2}}
		  \notag\\
		& = -\frac{\hbar^{2}}{2mr} \sqrt{\alpha_{1}\alpha_{2}}
		  \left(e^{-\alpha_{1}r} + e^{-\alpha_{2}r}\right).
		\label{eq:Wa}
\end{align}

To compare the two models, we use experimental parameters for the defects. The
binding energy of nitrogen is $|E_{2}| = 1.7$~eV, and that of the NV center is
$|E_{1}| = 2.6$~eV. For benchmarking, we adopt an electron effective mass of $m = 0.35,m_{e}$. Using $\alpha_{i} = \sqrt{2m|E_{i}|}/\hbar$, the
corresponding well parameters are:
\begin{align*}
\alpha_{1} & = 0.209\ a_{B}^{-1}, \\
\alpha_{2} & = 0.259\ a_{B}^{-1},
\end{align*}
where $a_{B}$ is the Bohr radius, corresponding to $0.529$~\AA.

Figure~\ref{fig:WWacomp} compares $W^{2}$ calculated using the full treatment,
which includes wave function overlaps as described by Eq.~\eqref{eq:Wfinal},
with the zero-overlap approximation given by Eq.~\eqref{eq:Wa}. The bottom panel
of the figure shows the ratio of the two models, indicating the differences
between them. We find that neglecting wave function overlaps results in a
significantly faster decay of $W^{2}$ with increasing distance. This
demonstrates the important role of wave function overlaps in providing an accurate
qualitative and quantitative description of the transfer integral.

\begin{figure}
\centering
\includegraphics[width=0.5\textwidth]{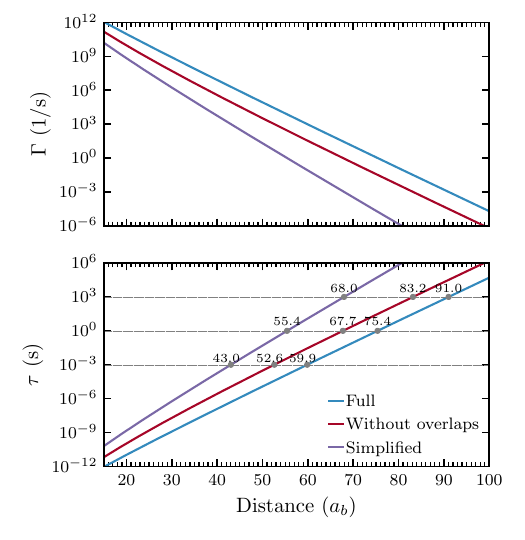}
\caption{Comparison of calculated rates and lifetimes using the full treatment
  (including overlaps) the zero-overlap approximation and single
  exponential.\label{fig:WWacomp}}
\end{figure}

\subsection{Benchmarking the Zero-Range Potential Model Against a Finite Spherical Well}

\begin{figure}
\centering
\includegraphics[width=0.5\textwidth]{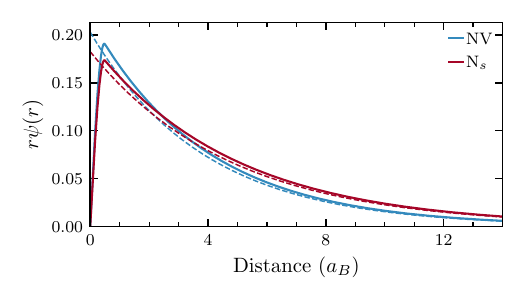}
\caption{Radial wavefunctions \(r\psi(r)\) for the ZRP and finite spherical well models. The dashed lines represent the ZRP wavefunctions, while the solid lines correspond to the finite spherical well model.\label{fig:wfs}}
\end{figure}

\begin{figure}
\centering
\includegraphics[width=0.5\textwidth]{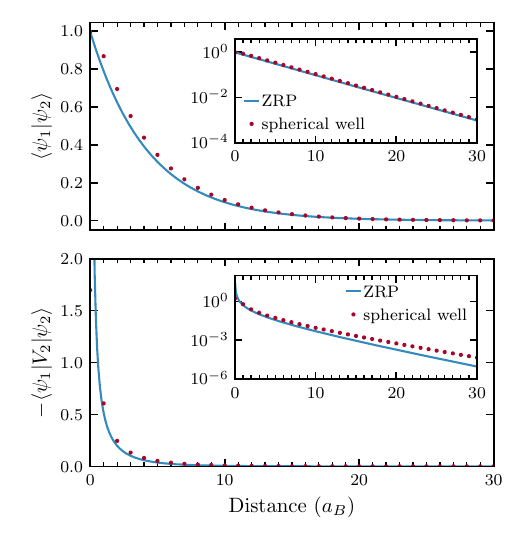}
\caption{Comparison of overlaps and potential term contributions \(-\langle \psi_{2} | V_{2} | \psi_{1} \rangle\) for the ZRP and finite spherical well models. The analytical expressions (Eqs.~\eqref{eq:ta1} and \eqref{eq:overlap}) are shown alongside the numerical results.\label{fig:overlaps}}
\end{figure}

While the zero-range potential (ZRP) model provides a simplified and
analytically tractable framework, it neglects the finite potential depth and
spatial extent, which can influence tunneling dynamics. To evaluate the accuracy
of the ZRP model in describing electron tunneling between two deep defects, we
benchmark it against a more realistic model involving finite spherical wells,
defined by the potential:
\[
V(r) =
\begin{cases}
-V_{0} & r \leqslant a, \\
0 & r > a,
\end{cases}
\]
where \(a\) is the radius of the potential well. In this study, we set
\(a = a_{B}/2\), where \(a_{B}\) is the Bohr radius. The depth of the potential
well, \(V_{0}\), is chosen to ensure that the system supports a single bound
state with an ionization energy matching those of the actual \(\NV\) and \(\Ns\)
defects. This yields \(V_{0} = 425.44~\text{eV}\) and
\(V_{0} = 417.203~\text{eV}\) for the \(\NV\) and \(\Ns\) centers, respectively.

\begin{figure}
\centering
\includegraphics[width=0.5\textwidth]{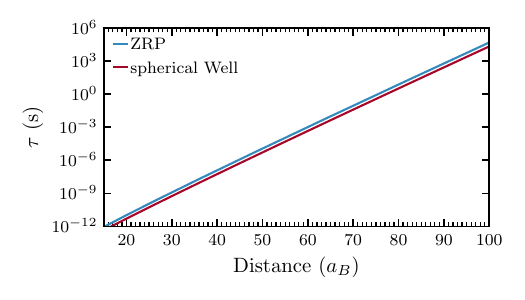}
\caption{Comparison of numerically calculated tunneling rates with the
  analytical expression of Eq.~\eqref{eq:Wfinal}.\label{fig:rates}}
\end{figure}

In Fig.~\ref{fig:wfs}, we compare the radial dependence of the spherical well
wavefunctions \(r\psi_{i}(r)\) with the corresponding ZRP wavefunctions,
\(c_{i} \exp(-\alpha_{i} r)\), shown as dashed lines. In
Fig.~\ref{fig:overlaps}, we present the numerically calculated overlaps and the
potential term contributions \(-\langle \psi_{2} | V_{2} | \psi_{1} \rangle\).
These results are compared with the analytical expressions provided in
Eqs.~\eqref{eq:ta1} and \eqref{eq:overlap}, demonstrating strong agreement.
Finally, in Fig.~\ref{fig:rates}, we compare the numerically calculated
tunneling rates with the analytical expression of Eq.~\eqref{eq:Wfinal}.

This analysis demonstrates that the ZRP model closely reproduces the tunneling
behavior of two narrow, finite-dimensional spherical wells, validating its
applicability as an effective approximation for electron transfer.

%% file: sup/pumping.tex
\section{Population buildup during ionization stage}

In the experiment, the ensemble is excited by 620\,nm laser pulses at a
repetition rate of 10~kHz (pulse spacing $\Delta t=100~\mu$s) with an average
power of approximately 100~$\mu$W. For the timescales considered here, the
sequence of optical pulses can be treated as a continuous‐wave (CW) excitation
characterized by an average ionization rate $G$. Ionization of $\NVm$ centers
during this stage produces a transient $\NV^0$ population $P$ that can be described
by a rate equation:
\begin{equation}
\frac{dP}{dt}=-\Gamma P+(1-P)G,\label{eq:s1}
\end{equation}
where $P$ is the instantaneous $\NV^0$ population, $\Gamma$ is the recombination
rate, and $G$ is the effective average ionization rate. The analytical solution
of Eq.~\eqref{eq:s1} with $P(t_0)=0$ is
\begin{equation}
P(t)=\frac{G}{\Gamma+G}
\Bigl(1-e^{-(\Gamma+G)(t-t_{0})}\Bigr).\label{eq:s2}
\end{equation}
The population immediately after pumping for a duration $\Delta t$ is
\begin{equation}
P(\tau;\Delta t)
=\frac{G}{\Gamma+G}\Bigl(1-e^{-(\Gamma+G)\Delta t}\Bigr)
\label{eq:s3}
\end{equation}
Equation~\eqref{eq:s3} defines the filter function $f(\tau;\Delta t)$, which
quantifies how efficiently a subensemble with lifetime $\tau=1/\Gamma$ builds up
population during the ionization stage.

Since the ionization process proceeds via two‐photon absorption, it is
relatively slow, and the resulting ionization rate $G$ can be assumed to be
small. In the weak‐excitation limit ($G\ll\Gamma$), this simplifies to
\begin{equation}
f(\tau;\Delta t)\approx
G\,\tau\bigl(1-e^{-\Delta t/\tau}\bigr).
\label{eq:s4}
\end{equation}

In this model, the transient $\NV^0$ population as a function of the distance
between the nitrogen and $\NV$ centers immediately after the pump laser is
switched off is given by
\[
w_{\mathrm{eff}}(r) = w(r)\,f(\tau[r];\Delta t),
\]
where $w(r)$ is the radial probability distribution of NV--N pairs, $\tau[r]$
is the corresponding tunneling lifetime, and $r$ denotes the separation
between the nitrogen and $\NV$ centers.


%% file: sup/ensemble.tex
\section{NV in the ensemble of N$_{s}$ defects}

Let us now examine an ensemble of $\NV$ centers interacting with an ensemble of
$\Ns$ defects. The concentration of $\Ns$ defects, denoted as $n$, is assumed to
be much higher than the concentration of $\NV$ centers. Therefore, at
thermodynamic equilibrium, nearly all $\Ns$ defects are in their neutral charge
state ($\Ns^{0}$), while the NV centers are in their negatively charged state
($\NVm$). Upon photoionization of an $\NVm$ center, an electron is excited to
the conduction band, leaving the NV center in its neutral charge state
($\NV^{0}$). Charge equilibration then proceeds via tunneling: the electron is
transferred from the nearest $\Ns^{0}$ donor to the photoionized NV center. This
restores the negative charge state of the NV and converts that specific donor to
$\Ns^{+}$. We assume that the positive charge formed on this nearest neighbor
does not last, since the ionizing excitation constantly redistributes the local
charge. Therefore, when estimating the tunneling rate, it is enough to assume
that all nitrogens are in their neutral charge state and to consider the
distance to the closest neutral nitrogen atom.

\subsection{Radial Probability of the Nearest Neighbor}

To estimate the probability of finding the nearest neighbor, we employ
a Poissonian model for the system. This approach is valid because
the probability of finding a nitrogen atom within a small volume element
is low, and the likelihood of finding two $\Ns$ defects in an infinitesimally
small volume $\mathrm{d}V$ is negligible. Furthermore, the occurrence
of one nitrogen atom is independent of the presence of other nitrogen
atoms. In such cases, the probability of finding $k$ nitrogen substitutionals
within a volume $V$ is given by the Poisson distribution:
\[
f(k,V)=\frac{\lambda^{k}e^{-\lambda}}{k!},
\]
where $\lambda=nV$ represents the expected number of nitrogen atoms
within the volume $V$, and $n$ is the concentration of nitrogen
atoms. In terms of the radial distance $r$ from the NV center, this
probability is given by:
\[
f(k,r)=\left(\frac{4}{3}\pi r^{3}n\right)^{k}\frac{e^{-\frac{4}{3}\pi r^{3}n}}{k!}.
\]

The probability of finding the nearest neighbor
in the shell between $r$ and $r+\mathrm{d}r$ is the product of two
factors: (1) the probability of finding zero nitrogen atoms within
a sphere of radius $r$, given by $f(0,r)$; and (2) the radial density
of nitrogen atoms at distance $r$,
expressed as $p(r)=4\pi r^{2}n$. Combining these factors, the radial
probability distribution for the nearest $\Ns$ defect around a given
NV center is:

\[
w(r)=4\pi r^{2}ne^{-\frac{4}{3}\pi nr^{3}}.
\]
The function $w(r)$ reaches its peak at a distance of $(2\pi n)^{-1/3}$.

\begin{figure}
\centering
\includegraphics[width=0.5\textwidth]{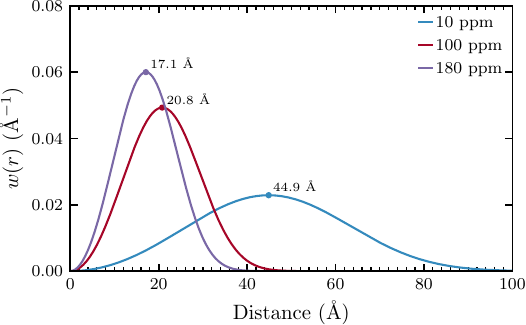}
\caption{Radial distribution of $\Ns$ defects for different concentrations. The
  distributions illustrate the likelihood of finding the nearest $\Ns$ defect at
  a distance $r$ from a given NV center.\label{fig:wrad}}
\end{figure}

In the experiment, the samples used have nitrogen concentrations of $n=10$ and $180$ ppm. As the atomic density of carbon in diamond is
$n_{C}=2.612\times10^{-2}~\mathrm{atoms}/a_{B}^{3}$, these nitrogen
concentrations correspond to $n=2.612\times10^{-7}$ and
$4.702\times10^{-7}~\mathrm{atoms}/a_{B}^{3}$, respectively. In
Fig.~\ref{fig:wrad}, we present the radial distribution of the nearest neighbors
corresponding to the densities mentioned above. As an approximation, we take the values of the distribution peaks as the
mean distance between $\Ns$ and NV center 44.9 and 17.1 \AA for 10 and 180 ppm concentrations respectively. 

%% file: sup/exp.tex
\section{Experimental details}

The 10 ppm sample is a CVD-grown diamond (Element Six) with a total NV concentration of 4.5ppm, of which about 3 ppm are $\NV^{-}$ and 1.5 ppm are $\NV^{0}$ in equilibrium. The 180 ppm sample is HPHT diamond with a $\NV^{-}$ concentration of 0.5 ppm.

\begin{figure}[b]
\centering
\includegraphics[width=0.85\textwidth]{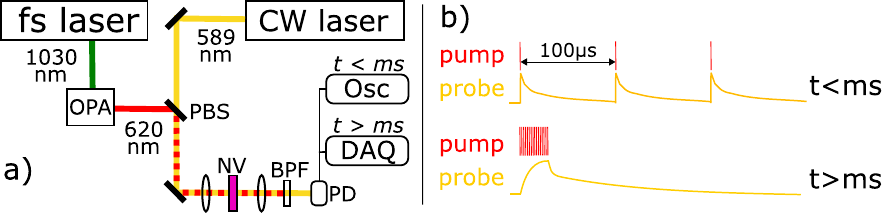}
\caption{a) Measurement setup, OPA: optical parametric amplifier, PBS: polarizing beam splitter, BPF: band-pass filter, PD: photodiode; b) excitation scheme for short-time (sub-ms) dynamics and long-time dynamics.\label{fig:setup_SI}}
\end{figure}

For the experimental investigation of charge equilibration, we employ a pump-probe
spectroscopy setup in which a pulsed pump laser impulsively excites an ensemble
of NV centers, while a continuous-wave (CW) laser probes the time-resolved
transmission through the sample. The setup and excitation schemes are
illustrated in Fig.~\ref{fig:setup_SI}(a). Unlike conventional pump-probe
configurations, where both the pump and probe are
pulsed~\cite{ulbricht2016,ulbricht2018a,carbery2024}, our approach employs a CW
probe laser. In pulsed setups, the time resolution is determined by the pulse
duration, but the scanning range is typically limited to around 10~ns, which is
insufficient for our study. By using a CW probe laser, we achieve longer time
scan ranges, extending up to seconds, with the time resolution determined by the
electronic bandwidth of the measurement acquisition system.\looseness=-1

The pump pulse with a pulse duration of about 150~fs at a wavelength of 620~nm
is generated in a homebuilt optical parametric amplifier (OPA) that is pumped by
the output of an Ytterbium-based fiber laser (Amplitude Tangerine SP) set at a
pulse repetition rate of 10~kHz. The wavelength is chosen to selectively excite
$\NVm$ but not $\NV^{0}$. The CW laser, attenuated to a power of about
100~\textmu W, emits at a wavelength of 589~nm, which coincides with the
absorption window of $\NV^{-}$ but not $\NV^{0}$. The pump power is set to 30 mW. Both beams are collinearly combined by a
polarizing beam splitter and focused to a spot size of 30~\textmu m on the
sample, which is housed in a closed-cycle Helium cryostat. An optical bandpass
filter placed after the cryostat rejects the pump light and any collected NV
photoluminescence (PL), transmitting only the probe beam, which is imaged on a
silicon photodiode (PD). The PD output voltage dynamics are digitized either
with a fast digital oscilloscope to resolve sub-ms-long dynamics with nanosecond
time resolution or with a data acquisition (DAQ) card to observe dynamics with
durations beyond the millisecond range. For the latter case, the pulsed pump
laser excites the sample for a duration of 200~ms, thus building up a
non-equilibrium state of ionized NV centers. After switching off the pump, the
system is allowed to relax back to equilibrium for up to ten seconds while
monitoring this process with the probe laser. This cycle is repeated several
times, and the recorded transient transmission data are averaged. In this
scenario, we also employ a balanced photodetection scheme to reduce long-term
probe laser fluctuations, where part of the probe beam is split off before the
sample and imaged on a reference photodiode.

\begin{figure}
\centering
\includegraphics[width=0.98\textwidth]{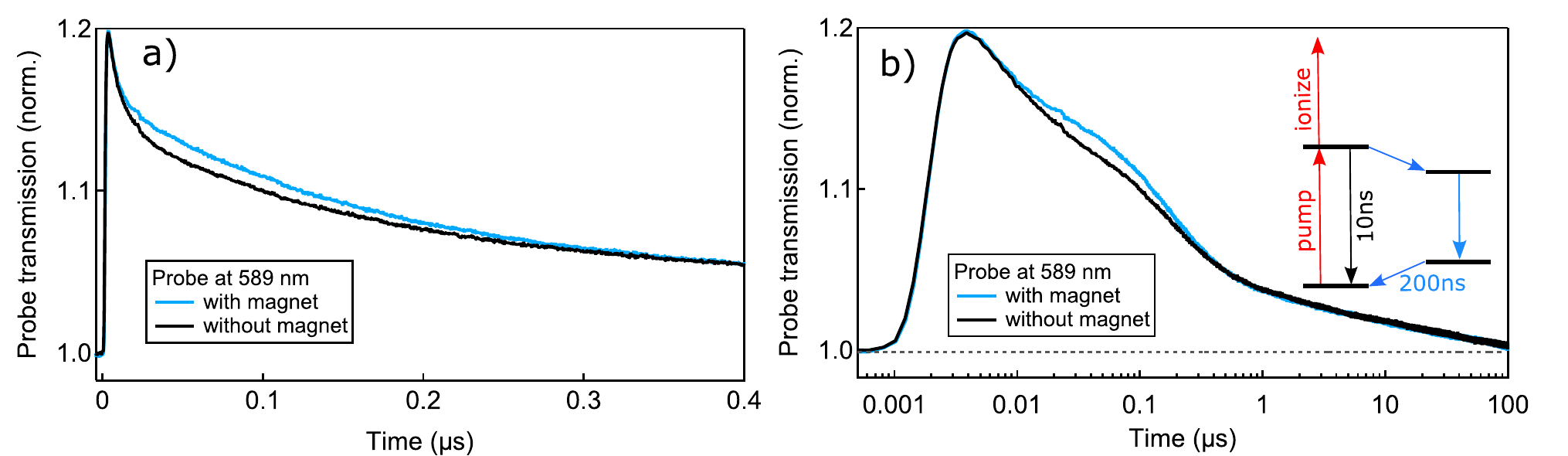}
\caption{Short-time transients of the probe transmission at 589 nm, with and without permanent magnet; a) up to 0.4 $\mu$s; b) up to 100 $\mu$s; the inset sketches the intrinsic $\NVm$\ electronic levels and occurring processes after photoexcitation. \label{fig:shorttime}}
\end{figure}

In order to elucidate the relaxation dynamics of photoexcited $\NVm$\ centers, we probed their ground-state bleaching (GSB) by monitoring the $\NVm$\ absorption at 589 nm. Fig.~\ref{fig:shorttime} shows the probe transmission transients between each pump pulse up to 100 $\mu$s after photoexcitation (defined by the 10 kHz repetition rate), normalized to the value before excitation. The relaxation in the first 400 ns can be fitted by a bi-exponential decay model, with a fast time constant of 9 ns and a slow time constant of about 200 ns. The fast component is due to the direct relaxation to the ground state via radiative recombination, indicated by the black arrow in the inset of Fig.~\ref{fig:shorttime}~b). The slow component is due to relaxation via intersystem crossing (ISC) through the singlet levels, sketched by the blue arrows \cite{doherty2013a,ulbricht2018a,robledo2011}. The relaxation through the ISC path is corroborated by a modification of the transient after placing a strong permanent magnetic close to the sample (blue plot). In that case, the spin polarization in the $\NVm$\ ground state achieved through periodic excitation is scrambled, thus forcing more $\NVm$\ centers to relax via the ISC path. This effect is similar to resonant microwave driving of the $m_s$ = 0 to $m_s$ = $\pm$1 transition.

If only dynamics intrinsic to $\NVm$\ were occurring, all photoexcited $\NVm$\ centers should have relaxed back to the ground state after 400 ns, with the GSB signal returning to its level before excitation, i.e. 1. This is clearly not the case because photoionized $\NVm$\ lead to the normalized probe transmission showing values $>$ 1 past 100 $\mu$s, which is when the next pump pulse arrives. As a consequence, the train of repetitive pump pulses also slowly builds up long-time GSB bleaching past the repetition time of the laser, thus acting more like a CW laser on these longer timescales. This scenario is used for the measurements featured in the manuscript, where we probe the bleaching dynamics on millisecond to second timescales, with the pump laser output being modulated by a mechanical shutter. When opened, the long-time GSB is slowly building, as evidenced by an increase in probe transmission. After 200 ms, the pump is blocked again and the system allowed to relax without perturbation, with the bleaching dynamics being monitored by the probe laser. This cycle is repeated several times, and the recorded transient transmission data are averaged.

\begin{figure}
  \centering
  \includegraphics[width=0.5\linewidth]{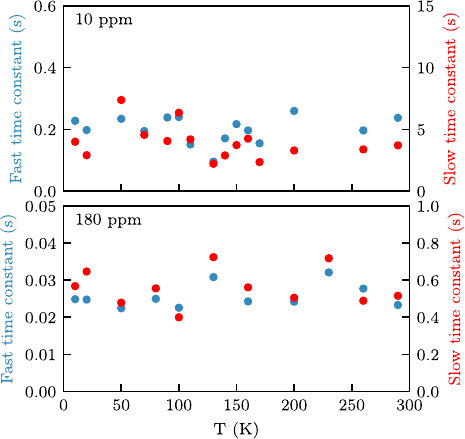}
  \caption{Temperature dependence of the two time constants from the bi-exponential fits to the recovery dynamics of the (a) 10 ppm and (b) 180 ppm sample.
    \label{fig:temp}}
\end{figure}

The data can generally be fitted well by a bi-exponential decay function, which provides a convenient way to quantitatively compare the dynamics at different temperatures. Fig. \ref{fig:temp} shows the extracted time constants for the 10 ppm sample (a) and the 180 ppm sample (b) for a temperature range from 10K to room temperature, where no discernible temperature dependence can be recognized.

